\documentclass[10pt,superscriptaddress,twocolumn,aps,prl]{revtex4-2}
\usepackage{amsmath}
\usepackage{amssymb}
\usepackage{graphicx}
\usepackage[colorlinks=true,citecolor=blue,linktocpage=true,linkcolor=blue,filecolor=blue,urlcolor=magenta]{hyperref}
\usepackage{physics}
\usepackage{natbib}
\usepackage{times}
\graphicspath{{./figures/}}
\usepackage{bbm}
\newcommand{\unit}[1]{\mathrm{#1}}
\newcommand{\beginsupplement}{%
            \setcounter{table}{0}
            \renewcommand{\thetable}{S\arabic{table}}%
            \setcounter{equation}{0}
            \renewcommand{\theequation}{S\arabic{equation}}%
            \setcounter{figure}{0}
            \renewcommand{\thefigure}{S\arabic{figure}}%
     }

\begin{document}
\title{Generating time-domain linear cluster state by recycling superconducting qubits}
\author{Shotaro Shirai}
\email{shirai-shotaro@g.ecc.u-tokyo.ac.jp}
\altaffiliation{Present address : Komaba Institute for Science (KIS), The University of Tokyo, Meguro-ku, Tokyo, 153-8902, Japan}
\affiliation{Department of Physics, Tokyo University of Science, 1--3 Kagurazaka, Shinjuku, Tokyo 162--0825, Japan}
\affiliation{RIKEN Center for Quantum Computing (RQC), 2--1 Hirosawa, Wako, Saitama 351--0198, Japan}
\author{Yu Zhou}
\affiliation{RIKEN Center for Quantum Computing (RQC), 2--1 Hirosawa, Wako, Saitama 351--0198, Japan}
\author{Keiichi Sakata}
\affiliation{Department of Physics, Tokyo University of Science, 1--3 Kagurazaka, Shinjuku, Tokyo 162--0825, Japan}
\author{Hiroto Mukai}
\affiliation{Department of Physics, Tokyo University of Science, 1--3 Kagurazaka, Shinjuku, Tokyo 162--0825, Japan}
\affiliation{RIKEN Center for Quantum Computing (RQC), 2--1 Hirosawa, Wako, Saitama 351--0198, Japan}
\author{Jaw-Shen Tsai}
\affiliation{Department of Physics, Tokyo University of Science, 1--3 Kagurazaka, Shinjuku, Tokyo 162--0825, Japan}
\affiliation{RIKEN Center for Quantum Computing (RQC), 2--1 Hirosawa, Wako, Saitama 351--0198, Japan}

\begin{abstract}
Cluster states, a type of highly entangled state, are essential resources for quantum information processing. Here we demonstrated the generation of a time-domain linear cluster state (t-LCS) using a superconducting quantum circuit consisting of only two transmon qubits. By recycling the physical qubits, the t-LCS equivalent up to four physical qubits was validated by quantum state tomography with a fidelity of 59$\%$.
We further confirmed the true generation of t-LCS by examining the expectation value of an entanglement witness. Our demonstrated protocol of t-LCS generation allows efficient use of physical qubits which could lead to resource-efficient execution of quantum circuits on a large-scale.
\end{abstract}

\maketitle
Practical quantum computing requires university, scalability, and fault tolerance. Although great advances towards intermediate-scale have been made recently in various systems, such as superconducting qubits \cite{arute2019}, trapped ions \cite{bruzewicz2019trapped} and semiconductor quantum dots \cite{kloeffel2013prospects}, scale-up of such systems while keeping addressing and manipulating each qubit individually is still experimentally challenging. Measurement-based quantum computation \cite{nielsen2006cluster} could be an alternative route, which is a scheme based on the generation of large-scale cluster states with a three-dimensional (3D) topology \cite{Raussendorf2007}.

As a type of genuine multipartite entangled (GME) states, cluster states \cite{Raussendorf2001} are generated in lattices of qubits with Ising-type interactions. To construct the 3D cluster state with qubits arranged only in the spatial domain, many difficulties would emerge, such as reduced accessibility to individual qubits and a cubic increase in footprints. While by converting one spatial dimension of the cluster into a temporal domain, the complexity of the 3D cluster state in space could be reduced to two \cite{Raussendorf2007, 3dcs}.

To date, 1D and 2D cluster states have been experimentally generated in several ways. In optics, both 1D and 2D cluster states on a large-scale have already been generated recently \cite{asavanant2019generation, larsen2019deterministic}. However, for fault tolerance, it is necessary but difficult to encode a qubit into the phase space \cite{GKP2001, Bourassa2021blueprintscalable}. While in superconducting quantum circuits, only linear (1D) cluster states (LCSs) are experimentally demonstrated by using either a 1D array of transmons \cite{gong2019genuine} or microwave photonic qubits \cite{besse2020realizing, ilves2020demand}. In the former experiment, the size of the LCS is determined by the number of physical qubits. The latter one relies on microwave photons to generate the entangled states, which also suffer from the loss of photons in the waveguides. The states generated by the above methods, when not considering the measurement, are spatial domain cluster states because they exist in either an array of physical qubits or a train of propagating microwave photons.

In this work, we demonstrated the generation of a time-domain LCS (t-LCS) using a superconducting quantum circuit with only two physical qubits. By recycling one of the qubits, we could generate a four-qubit t-LCS with a fidelity of 59$\%$. We further confirmed the four-qubit t-LCS is the GME state by examining the expectation value of the projector witness \cite{guhne2009entanglement}. Our demonstrated protocol for t-LCS generation shows the feasibility of generating entangled states with a size larger than the number of physically available stationary qubits, which could finally lead to the 3D cluster generations with reduced complexity in the spatial domain and the resource-efficient implementations of large-scale quantum circuits.
\begin{figure}
    \centering
    \includegraphics[width=8.6cm]{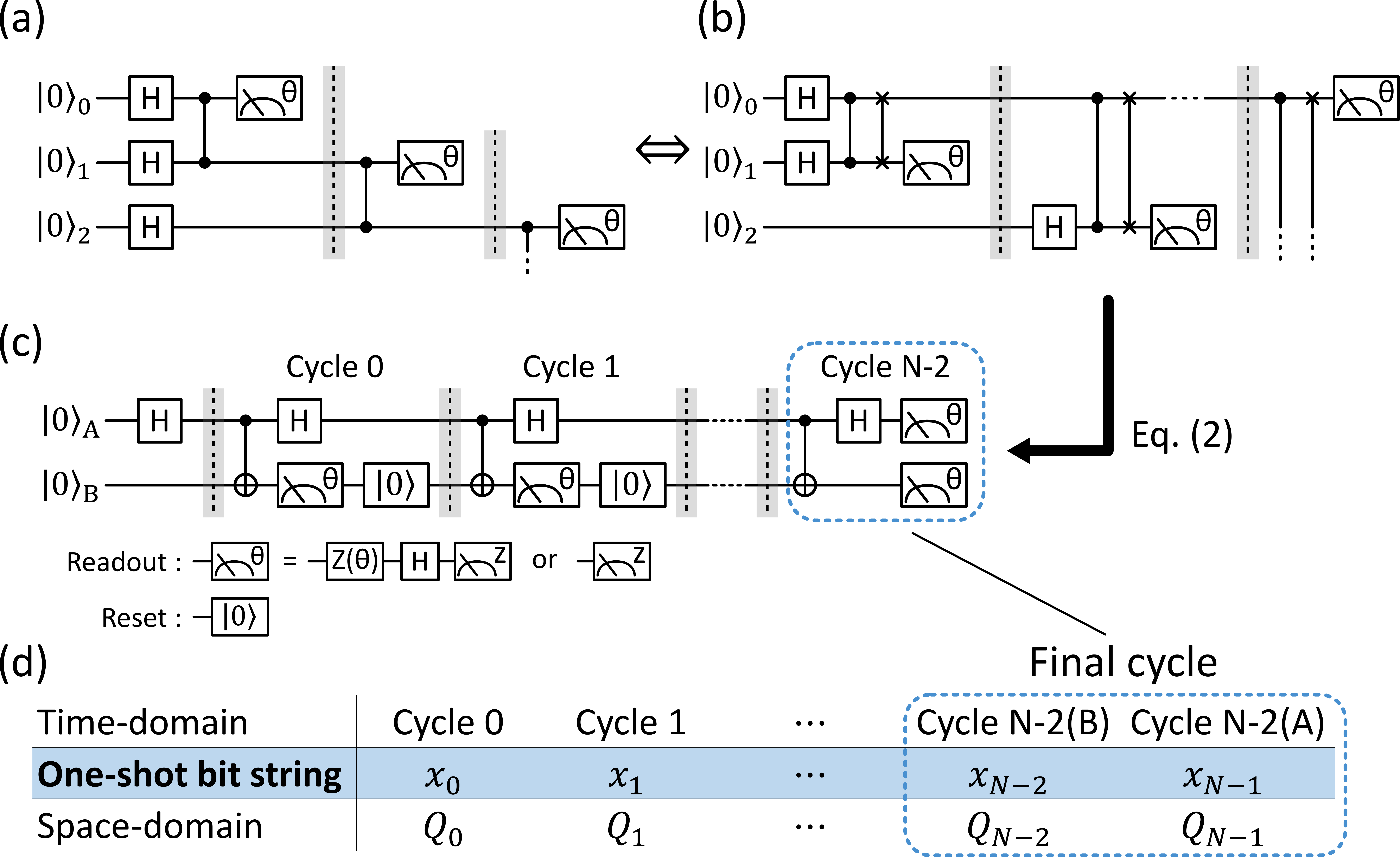}
    \caption{(a) Sequential quantum circuits that generate an LCS in the spatial domain. (b) Quantum circuit equivalent to (a) in which the measurement order is shifted by swap (SWAP) gate. (c) Quantum circuit for generating a t-LCS corresponding to $N$ qubit. The cycle (between dashed lines) is repeated $N-1$ times, and in the final cycle, both qubit A and qubit B are measured. The measurement basis is transformed using a Hadamard gate and an arbitrary phase rotation $Z(\theta)$. (d) Correspondence between measurements in the space and time domains.}
    \label{fig:s-t_LCS_equal}
\end{figure}

\emph{Generation protocol for t-LCS}---The LCS in the spatial domain is prepared by the following procedure [Fig.~\ref{fig:s-t_LCS_equal}(a)]. First, a state $\ket{+}^{\otimes N}$ is prepared by applying a Hadamard gate (H) to all $N$ qubits initialized to $\ket{0}$. Then, a controlled-phase (CZ) gate ($U_\mathrm{CZ}^{i,i+1}$) is applied between the $(i)$-th and $(i+1)$-th qubit ($i=0,1,...,N-1$). This state is represented by 
\begin{equation}
    \ket{\mathrm{LCS}_N} = \left[ \prod_{i=0}^{N-1} U_\mathrm{CZ}^{i,i+1} \right] H^{\otimes N}\ket{0}^{\otimes N}.
    \label{eq:LCS_N}
\end{equation}
The LCS can simulate the gate operation for one qubit by following the procedure of the measurement-based quantum computation \cite{nielsen2006cluster}. In this state, namely, a graph state \cite{nielsen2006cluster, guhne2009entanglement}, qubits correspond to nodes, and the CZ gates correspond to edges, which connect the nodes. To prepare this state, the operation $U_\mathrm{CZ}^{i+1,i+2}$ connecting the $(i+1)$-th and $(i+2)$-th nodes can be delayed for the measurement of the $(i)$-th qubit, as shown in Fig.~\ref{fig:s-t_LCS_equal}(a) \cite{nielsen2006cluster}. Here, we can rearrange the quantum circuit in Fig.~\ref{fig:s-t_LCS_equal}(a) as the circuit shown in Fig.~\ref{fig:s-t_LCS_equal}(b) using a SWAP gate. Then, the original 0-th measured state is transferred to the first qubit by using the SWAP gate, and the measurement is performed. Note that in Fig.~\ref{fig:s-t_LCS_equal}(b), the edges of the cluster state are always connected through the 0-th qubit. This allows us to reduce the number of physical qubits by combining the on-demand reset of the first qubit and the following identity
\begin{equation}
    U_\mathrm{CZ} \qty(I\otimes H) \ket{\psi}_\mathrm{A}\ket{0}_\mathrm{B} =  U_\mathrm{SWAP} \qty(H \otimes I) U_\mathrm{CNOT} \ket{\psi}_\mathrm{A}\ket{0}_\mathrm{B}. \label{eq:SWAP_elimination}
\end{equation}
Here, the labels $Q_0$ and $Q_1$ have been changed to $Q_\unit{A}$ and $Q_\unit{B}$ in Fig.~\ref{fig:s-t_LCS_equal}(c), respectively and a state $\ket{\psi}_A\otimes \ket{0}_B$ in Eq.~\eqref{eq:SWAP_elimination} is a input state for the $i$-th cycle prepared by the reset of $Q_\unit{B}$ at the end of the $(i-1)$-th cycle. Finally, using $U_\unit{SWAP}U_\unit{SWAP}=I$, the quantum circuit shown in Fig.~\ref{fig:s-t_LCS_equal}(c) is derived, and the t-LCS can be generated by executing this circuit.

When the quantum circuit of Fig.~\ref{fig:s-t_LCS_equal}(a) is executed once and a projection measurement is performed on qubit $Q_0$ - $Q_{N-1}$, a classical bit string $(x_0, x_1, ..., x_{N-1}),\,x_i \in \{0,1\}$ is obtained as shown in the blue row of Fig.~\ref{fig:s-t_LCS_equal}(d). This means that a LCS is projected to $\ket{x_0, x_1, ..., x_{N-1}}$. To generate a LCS corresponding to $N$ qubits in the time domain, the measurement and reset cycle can be repeated $N-1$ times. At this time, the measurement results of the $k=0,...,N-2$ cycles for $Q_\unit{B}$ correspond to the measurement results of the $(x_0, x_1, ..., x_{N-2})$ in Fig.~\ref{fig:s-t_LCS_equal}(d), and the $(N-1)$-th measurement result $x_{N-1}$ is obtained by measuring $Q_{A}$ in the end of total sequence. In this work, we generate the t-LCS by following the above protocol [Fig.~\ref{fig:s-t_LCS_equal}(c)].

\emph{Device and experimental setup}---We use a superconducting quantum circuit with a general layout consisting of two capacitively coupled fixed-frequency Xmons \cite{Charge-insensitive, Xmon}, two $\mathrm{\lambda}/4$-CPW readout resonators, and a Purcell filter \cite{reed2010fast, sank2014fast}. For qubit $Q_\unit{A(B)}$, when the system is in the ground state, $\omega_{ge}^\unit{A(B)}/2\pi\simeq5.213\,(5.351)\,\unit{GHz}$ and the anharmonicity is  $\alpha^\unit{A(B)}/2\pi\simeq-301\,(-296)\,\unit{MHz}$. For the residual ZZ interaction, the frequency of $Q_\unit{B}$ is measured using a Ramsey interferometer with $Q_\unit{A}$ prepared in the ground state and the first excited state. The result is $\chi_\unit{ZZ}/2\pi\sim 0.51\,\unit{MHz}$. The coupling constant between $Q_\unit{A}$ and $Q_\unit{B}$ is calculated to be 7.7 MHz from $\chi_\unit{ZZ}$. For the readout resonator $R_\unit{A(B)}$, the resonant frequency is $\omega_r^\unit{A(B)}/2\pi\simeq6.746\,(6.682)\,\unit{GHz}$ and the dispersive shift is $\chi_d^\unit{A(B)}/2\pi\sim2.0\,(2.5)\,\unit{MHz}$. The coupling quality factor ($Q_\unit{coup}$) between the readout resonator and the external circuit is calculated from 
the FWHM to be $Q_\unit{coup}^\unit{A(B)}\sim470\,(350)$.
\begin{figure}[t]
    \centering
    \includegraphics[width=8.6cm]{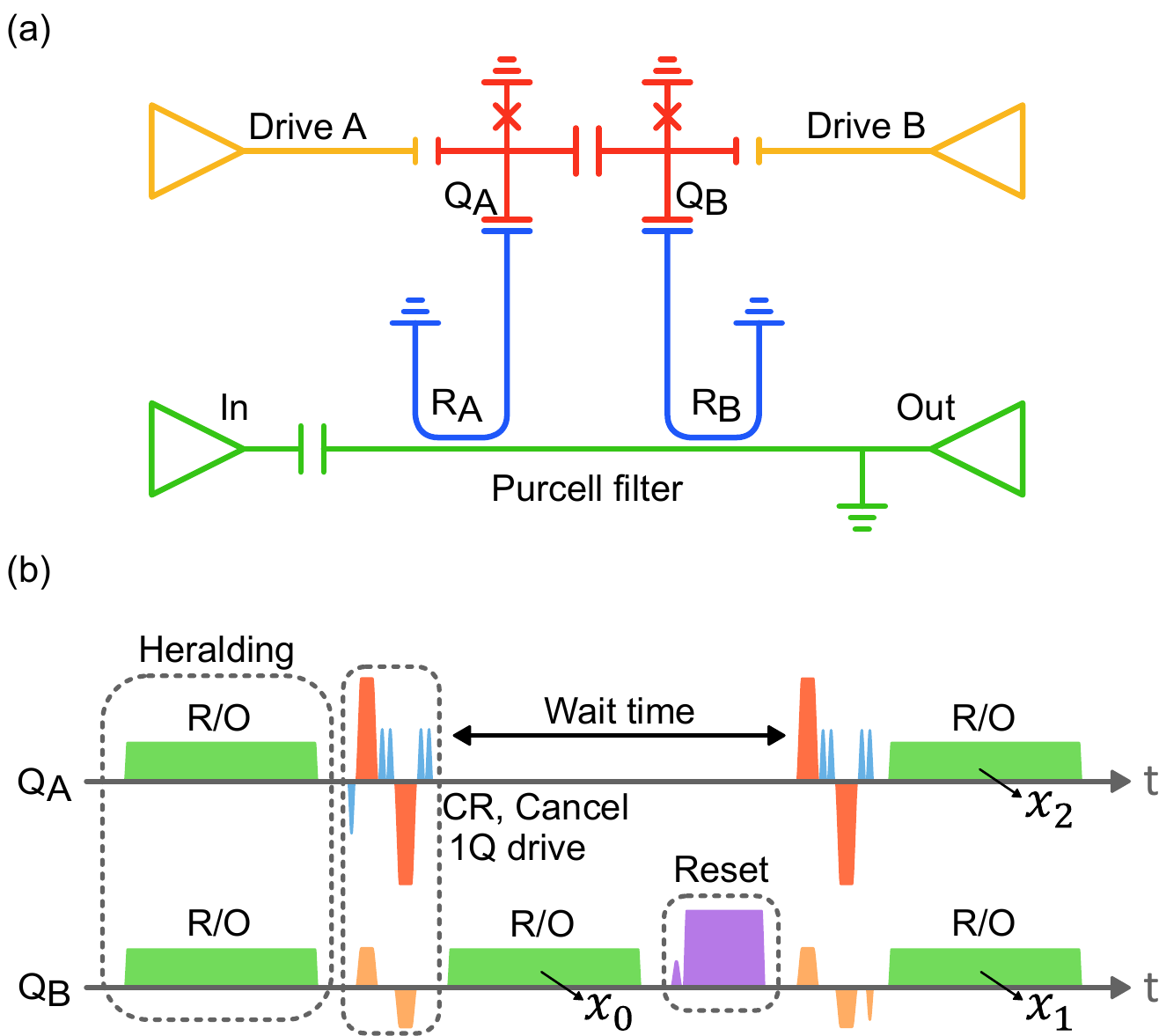}
    \caption{(a) Circuit diagram of the superconducting quantum circuit used for this work. The two transmons are directly capacitively coupled and the readout resonators are connected to a common Purcell filter. The local drive of qubit A and the CR drive are inputted from driveline A. The local drive of qubit B, the drive canceling the crosstalk of the CR drive, and the drive for resetting are inputted from driveline B. (b) Example of a pulse sequence for generating three-qubit t-LCS. To reduce the initialization error,  the initial state of each qubit is measured before the cycle begins.}
    \label{fig:pulseSeq_sample_schematics}
\end{figure}

Single-qubit gates are implemented using Gaussian pulses of $20\,\unit{ns}$ width combined with derivative removal by adiabatic
gate (DRAG) modulation \cite{DRAG_leak}, with phase correction by two Virtual-Z gates (VZ) \cite{mckay2017efficient}. The controlled-NOT (CNOT) gate uses an echoed cross-resonance (ECR) gate as a source gate \cite{Procedure2016Sheldon}, the pulse waveform is a flat-top Gaussian, and the CNOT gate time is 296 ns. These gates are optimized by optimized randomized benchmarking for immediate tune-up (ORBIT) \cite{ORBIT} after the initial value is estimated by error-amplifying experiments \cite{reed2013entanglement} and see Supplemental Material for ECR gate. The rectangular readout pulse length is 620 ns, and the signal is amplified by Josephson traveling wave parametric amplifier (JTWPA) \cite{macklin2015near}, also the integration time window length for the state classification is 750 ns. For the reset of $Q_\unit{B}$, we use the pulsed-reset protocol \cite{Pulsed_reset_protocol2018,PhysRevLett.121.060502} that can initialize the transmon qubit independently of whether its initial state is $\ket{0}$ or $\ket{1}$. The fidelity of the CNOT gate ($F_\unit{CNOT}$) is estimated by interleaved randomized benchmarking \cite{magesan2012efficient}, and the measured value is $F_\unit{CNOT}=0.967(12)$. Other operation fidelities and coherence times are summarized in Tab.~\ref{tab:sum_fidelity}.

\begin{table}[t]
    \centering
    \begin{tabular}{c@{\hspace{1em}}c@{\hspace{1em}}c@{\hspace{1em}}c@{\hspace{1em}}c@{\hspace{1em}}c}
            & $T_1\,\unit{[\mu s]}$ & $T_2^*\,\unit{[\mu s]}$ & $F_\unit{AGF}$ & $F_a$ & $F_{\ket{0}}$ \\ \hline \hline
    Qubit A & 20 & 29 & 0.9985(2)  & 0.950(4) & - \\
    Qubit B & 20 & 27 & 0.9993(1)  & 0.947(5) & 0.977(4) \\ \hline 
    \end{tabular}
    \caption{Typical coherence times of the sample and operation fidelities. $F_\unit{AGF}$ is the single qubit averaged gate fidelity. $F_a$ is the assignment fidelity of the single-shot readout. $F_{\ket{0}}$ is the reset fidelity. The measurement result for the reset fidelity is shown in the Supplemental Material. The numbers in brackets represent $95\%$ confidence.}
    \label{tab:sum_fidelity}
\end{table}

\begin{figure*}[t]
    \centering
    \includegraphics[width=17.8cm]{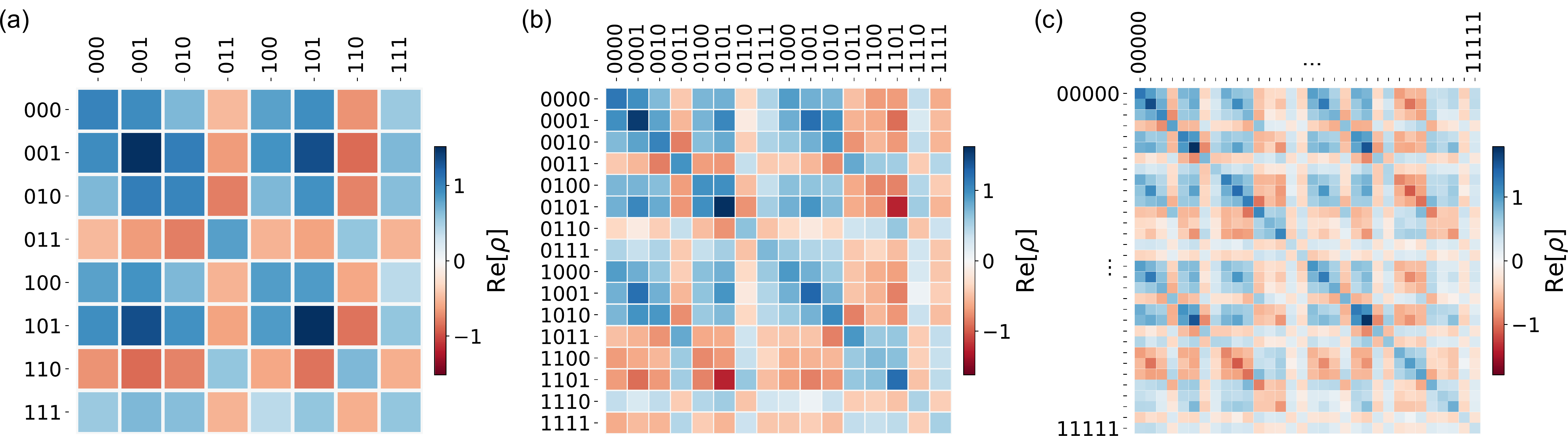}
    \caption{Real part of the reconstructed density matrix for the best state fidelity: (a) $F_3\simeq75\%$, (b) $F_4\simeq59\%$, and (c) $F_5\simeq47\%$. The matrix elements of the ideal LCS are all real values, and all absolute values are $1/d$, where $d$ is the dimension of the Hilbert space. In this plot, each matrix element is normalized by $1/d$.}
    \label{fig:LCS_cycle4_best75_59}
\end{figure*}

\begin{figure}[t]
    \centering
    \includegraphics[width=8.6cm]{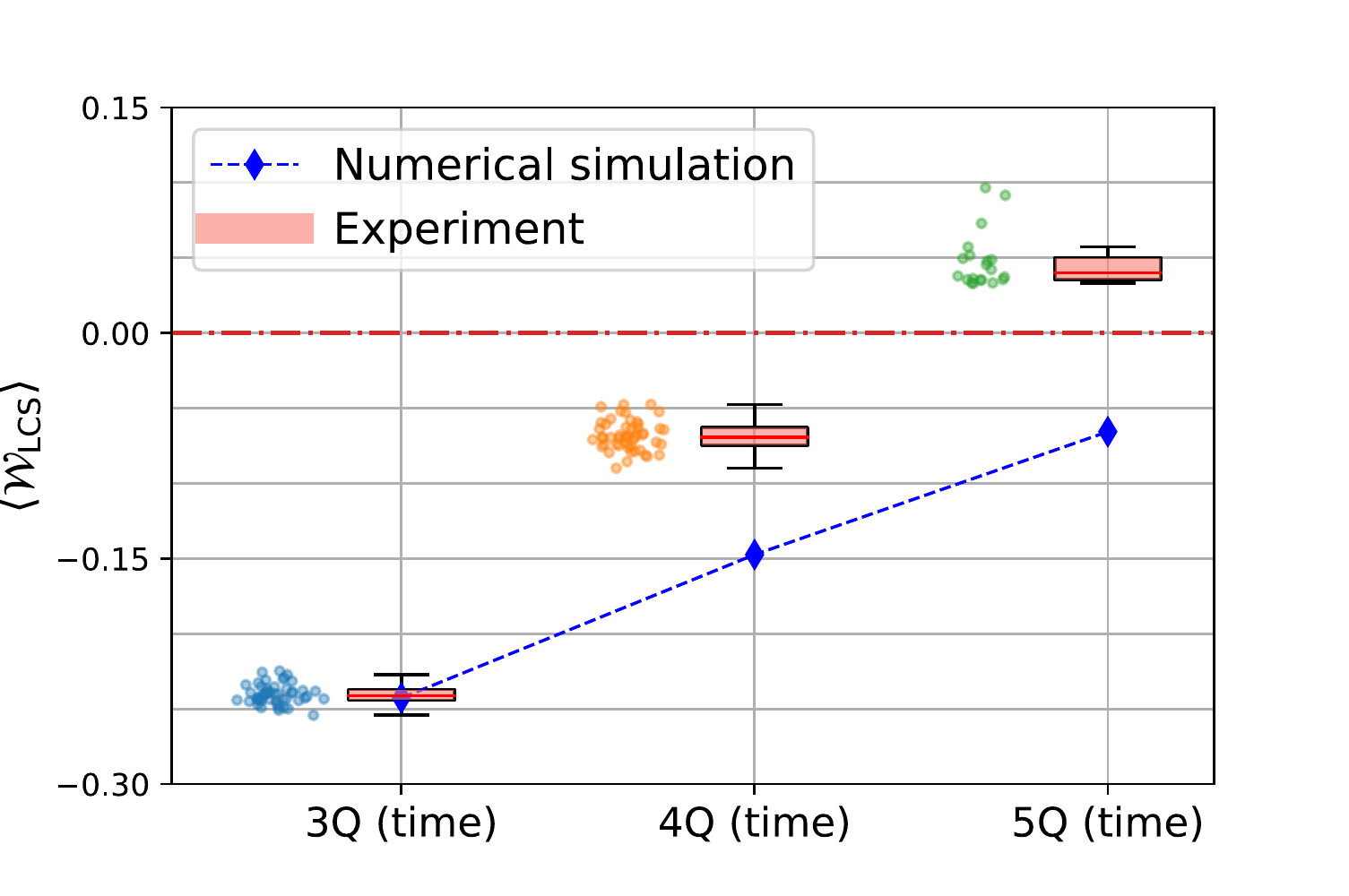}
    \caption{Projector witness values (scatter and box plot) were calculated from the results of quantum state tomography and their box-and-whisker plots. The number of measurement points was $50, 50,$ and $20$ for $N = 3, 4, $ and $ 5 $, and the measurement took 1.5 days in total. The blue dashed line shows the result of the numerical simulation, for which the standard deviations are omitted because they are about 10 times smaller than the experimental ones. The numerical simulation method is described in Supplemental Material.}
    \label{fig:Witness_cycle4}
\end{figure}

An example of a pulse sequence for executing the quantum circuit in Fig.~\ref{fig:s-t_LCS_equal}(c) is shown in Fig.~\ref{fig:pulseSeq_sample_schematics}(b). This circuit generates a three-qubit t-LCS. In the measurement, the initial state of each physical qubit is heralded before the t-LCS generation pulse to reduce the state preparation error. In the subsequent data processing, the expected value is calculated only when the heralding result is $\ket{00}$.
The purple pulse in Fig.~\ref{fig:pulseSeq_sample_schematics}(b) is a reset pulse composed of $X_{\pi}^{ef}$ and $X_{\pi}^{f0g1}$. The $X_{\pi}^{f0g1}$ pulse used to induce the interaction between the transmon and the resonator requires a large power and the crosstalk between the transmons causes a large AC Stark shift of $Q_\unit{A}$ and the extra phase shift. This phase shift of $Q_\unit{A}$ due to the AC-stark shift is corrected by $\unit{VZ}(\theta_\unit{z})$. See Supplemental Material for the determination of the amount of rotation $\theta_\unit{z}$.

\emph{Quantum tomography and Entanglement witness}---We performed quantum state tomography to characterize the generated states. First, quantum state tomography of the spatial-domain $N$-qubit system is outlined. The mixed state of the $N$-qubit system is expressed as
\begin{equation}
    \rho_N = \frac{\mathbbm{1}}{d} + \frac{1}{d}\sum_{n=1}^{d^N-1} p_n^{\boldsymbol{r}_n}\sigma_0^{r_n^0}\otimes\sigma_1^{r_n^1}\otimes\cdots\otimes\sigma_{N-1}^{r_n^{N-1}}.
\end{equation}
Here, $d$ is the dimension of the system, $p_n^{\boldsymbol{r}_n}=\Tr[\rho_N \sigma_0^{r_n^0}\otimes\sigma_1^{r_n^1}\otimes\cdots\otimes\sigma_{N-1}^{r_n^{N-1}}]$, $\sigma_i^{r_n^i}$ is one of the Pauli basis of the $i$-th qubit, and $\boldsymbol{r}_n=(r_n^0,r_n^1,...,r_n^{N-1}),\,r_n^i\in\{ \mathrm{I,\,X,\,Y,\,Z} \}$. The dispersive readout corresponds to the projection measurement in the Z basis. Measurements in other bases can be performed by applying the Z gate ($Z(\theta)=\exp[-i\theta/2\sigma^\unit{Z}]$) prior to the measurements as shown in the inset of Fig.~\ref{fig:s-t_LCS_equal}(c) and the measurement basis for each qubit was specified by $\boldsymbol{r}_n$.

Quantum state tomography for the t-LCS can be performed with a simple transformation from the above method. The measurement basis of the $i$-th qubit specified by $r_n^i$ corresponds to that in the $i$-th cycle except the final $(N-2)$-th cycle. In the $(N-2)$-th cycle, the measurement basis specified by $r_n^{N-2}$ and $r_n^{N-1}$ are performed on $Q_\unit{B}$ and $Q_\unit{A}$, respectively.

We performed quantum state tomography for up to four cycles, which simulates a five-qubit LCS. Each expectation value measurement consisted of $2 \times 10^3$ single-shot bit strings. After the heralding state preparation \cite{Heralded2012Johnson}, approximately $90\%$ of the data was used to calculate the expectation value because each qubit has a residual excitation of about $3\%$ in the steady state. The state fidelities $N=3,4$, and $5$ of the t-LCS is $F_3=74(1)\%,\,F_4=57(2)\%$, and $F_5=45(4)\%$, respectively, from the density matrix reconstructed using the least-squares method with the assumption of being a physical density matrix \cite{gross2010quantum}. The real part of the density matrices are shown in Fig.~\ref{fig:LCS_cycle4_best75_59}.

Next, the projector witness is used to verify that the generated state is a genuine multipartite entangled state. The projector witness for the LCS is defined by 
\begin{equation}
    \mathcal{W}_\unit{LCS} = \frac{\mathbbm{1}}{2} - \dyad{\unit{LCS}_N}{\unit{LCS}_N}.
    \label{eq:proj_witness}
\end{equation}
The expectation value of the projector witness is positive for all separable states, $\expval{\mathcal{W}_\unit{LCS}} < 0$ is obtained only for a state belonging to the GME state class, and the generated state has been determined to belong to the GME state \cite{guhne2009entanglement}. From Eq.~\ref{eq:proj_witness}, the relationship between the expectation value of this witness and the state fidelity is $\expval{\mathcal{W}_\unit{LCS}}=1/2-F_N$. Fig.~\ref{fig:Witness_cycle4} shows the projector witness values obtained by the result of quantum state tomography for 2,3, and 4 repetitions ($N=3,4,$ and 5 qubits in time-domain). It shows that a state with up to three repetitions, namely four-qubit t-LCS, is the GME state. 

Moreover, we also carried out the numerical simulation of the quantum circuit in Fig.~\ref{fig:s-t_LCS_equal}(c), as shown in Fig.~\ref{fig:Witness_cycle4}. We used the amplitude and the phase damping error channel \cite{geller2013efficient} to model the gate operation error. For the readout and reset error, a stochastic error model, using the infidelity of the assignment and reset as the error probability, was used. See Supplemental Material for the detailed definitions of the error channels. The numerical simulation quantitatively reproduced the experimental results. This confirms that our experimental protocol works well. In addition, although the size of the LCS generated in this work is limited by the exponential increase in the number of experiments required for quantum state tomography, by utilizing the fact that the LCS is a matrix product state, the experimental cost can be reduced to a linear increase with respect to the system size of $N$ \cite{PhysRevLett.111.020401}, and a larger system can be evaluated.

\emph{Conclusion}---We have demonstrated a protocol that can generate cluster states entangled in the time domain with very limited physical qubits. The fidelity of such t-LCSs is characterized by quantum state tomography, with 59$\%$ for the four-qubit equivalent case. We further confirm that such four-qubit t-LCS is a GME state from the expectation value of the projector witness. Our protocol can generate a t-LCS with a size larger than the number of available stationary physical qubits. Our demonstration also shows that, by further integration, it would be possible to generate a large-scale 3D cluster state by combining the current scheme with a qubit array arranged in a 2D lattice \cite{Raussendorf2007, 3dcs}. In addition, the demonstrated creation of larger entangled states by recycling the smaller number of physical qubits leads to resource-efficient implementations of quantum circuits on a large-scale \cite{Huggins_2019, PhysRevResearch.1.023025}.

\emph{Acknowledgement}---We acknowledge R. Wang, A. Tomonaga, K. Nittoh, K. Kusuyama and L. Szikszai for fabrication. We also acknowledge S. Kwon and K. Heya for helpful discussions and Lincoln Laboratory for providing the JTWPA. This work was supported by CREST, JST. (Grant No. JPMJCR1676). This work was supported by JST [Moonshot R\&D][Grant Number JPMJMS2067]. This paper was partly based on results obtained from a project, JPNP16007, commissioned by the New Energy and Industrial Technology Development Organization (NEDO), Japan.

\emph{Data availability}---The experimental data generated and analyzed during this work are available from the corresponding author on reasonable request.


\bibliographystyle{apsrev4-2}
\bibliography{bibs2}

\clearpage
\appendix
\onecolumngrid
\section{Supplemental Material for ``Generating time-domain linear cluster state by recycling superconducting qubits"} 

\beginsupplement
\section{Fabrication and experimental setup}
\begin{figure*}[b]
    \centering
    \includegraphics[width=13cm]{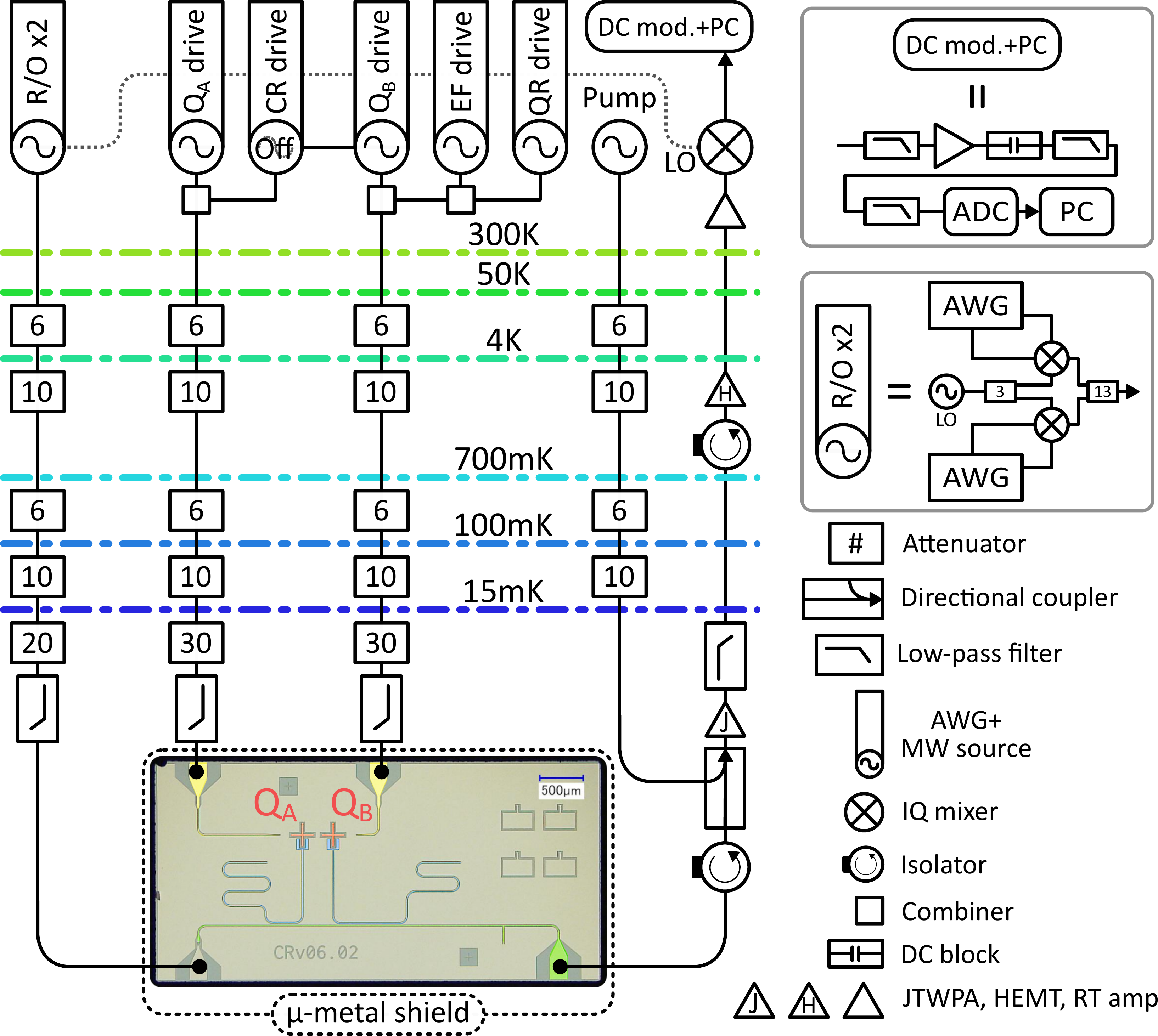}
    \caption{Setup of the experimental equipment.}
    \label{fig:setup_paper}
\end{figure*}

Our sample was fabricated on a $2.5\,\unit{mm}\times 5.0\,\unit{mm}$ high-resistivity Si substrate. Circuits such as resonators and waveguides were fabricated on 50-nm-thick Nb film by photolithography and reactive ion etching. The Josephson junctions were fabricated using the standard double-angle Al evaporation technique. The chip was protected by 3-layer magnetic shielding cooled to a base temperature of $\sim$15 mK. Control microwave pulses were produced by single sideband modulation (SSB), and the attenuation of each input line is about 60 dB to minimize the excitation of qubits by room-temperature black-body radiation. The cross-resonance (CR) drive and $Q_\unit{B}$ share the same local oscillator (LO) for phase coherence as shown in Fig.~\ref{fig:setup_paper}. The readout pulse for each qubit also shares the LO. The output signal was first amplified by the JTWPA and followed by a low-noise HEMT at 4K. At room temperature, the output signal was further amplified by a room temperature amplifier and then down-converted to an intermediate frequency and finally sampled by an analog-to-digital converter.

\section{Gate calibration}
\begin{figure}
    \centering
    \includegraphics[width=8.6cm]{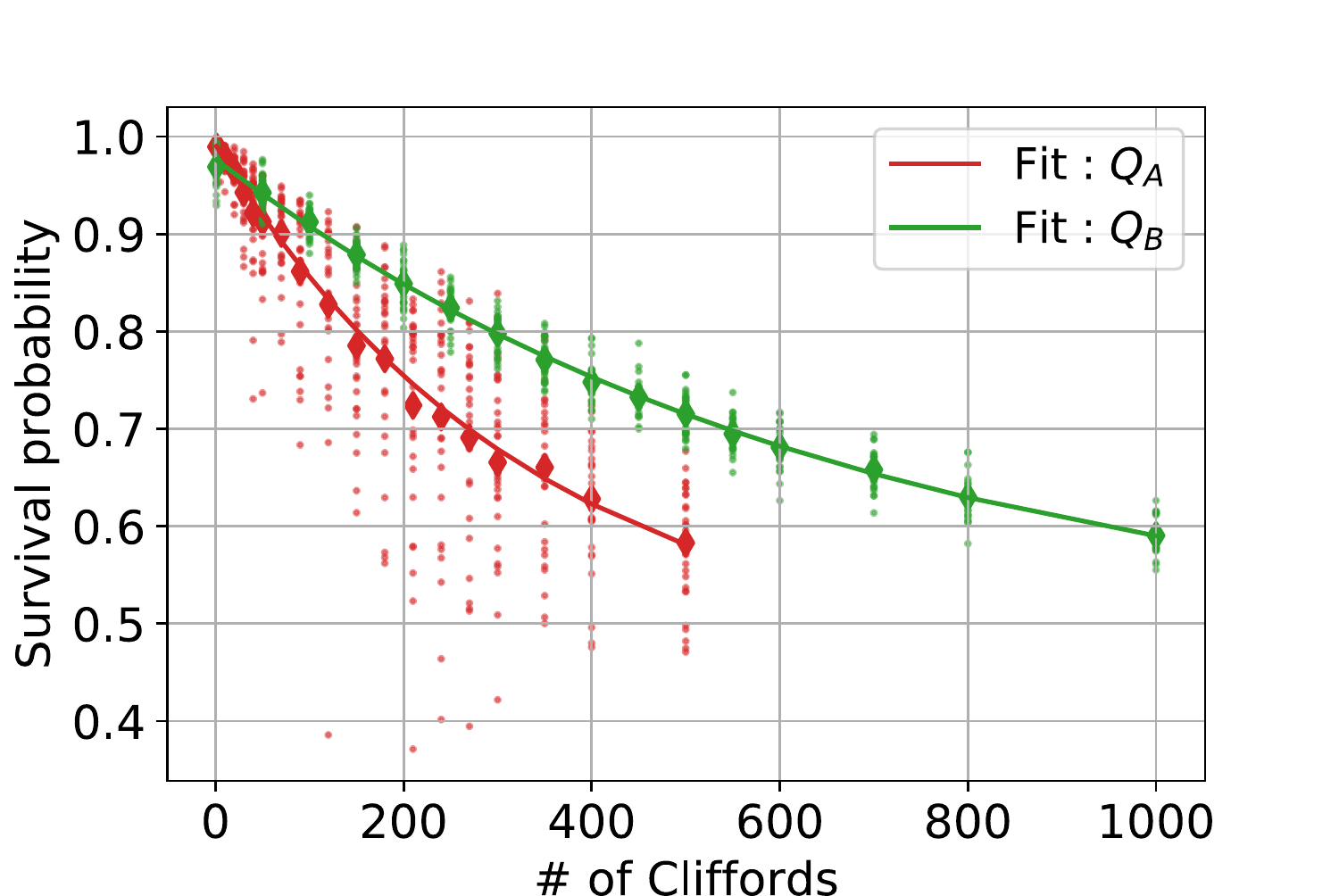}
    \caption{Result of single-qubit gate standard randomized benchmarking (SRB) after the optimization by ORBIT. For $Q_\unit{A}$, $F_\unit{AGF}=0.9985(2)$. For $Q_\unit{B}$, $F_\unit{AGF}=0.9993(1)$. The total number of seeds for each $m$-length Clifford gates is 30 and each point is the expectation value calculated from $10^4$ shots.}
    \label{fig:QAQB_SRB}
\end{figure}

The single-qubit gate used in this experiment was implemented by combining a $X_{\pi/2}$ gate as a 20 ns DRAG-modulated microwave pulse and two VZ gates. The local drive frequencies for $Q_\unit{A}$ and $Q_\unit{B}$ were $\omega_{d}^\unit{A}/2\pi\simeq\omega_{ge}^\unit{A}/2\pi\simeq5.210\,\unit{GHz}$ and $\omega_{d}^\unit{B}/2\pi\simeq\omega_{ge}^\unit{B}/2\pi+\chi_\unit{ZZ}/4\pi\simeq5.3513\,\unit{GHz}$, respectively. These frequencies were used to calibrate the amplitude error using an error amplification technique \cite{reed2013entanglement} to obtain a suitable initial pulse amplitude. Then, the pulse was optimized by ORBIT by sandwiching $X_{\pi/2}$ between the VZ gates with rotation angles $\theta_0^{pre},\, \theta_0^{pos}$ to correct the phase error \cite{mckay2017efficient} and the results to check the calibration performance is shown in Fig.~\ref{fig:QAQB_SRB}.  

The CNOT gate was optimized by ORBIT after initial values were obtained using the XY4 sequence \cite{ahmed2013robustness,garion2021experimental}. As the implementation method of the CNOT gate, we used Echoed cross-resonance (ECR) gate which is the same as the CNOT gate except for the single-qubit gates. Here, $Q_\unit{A}$ was a control bit and $Q_\unit{B}$ was a target bit. In our chip, the isolation between drivelines was 12 dB, so there was large crosstalk. To suppress this crosstalk, a method of simultaneously applying a cancel (CL) drive to $Q\unit{B}$ was used \cite{Reducing2020Sundaresan}.

\begin{figure}
    \centering
    \includegraphics[width=8.6cm]{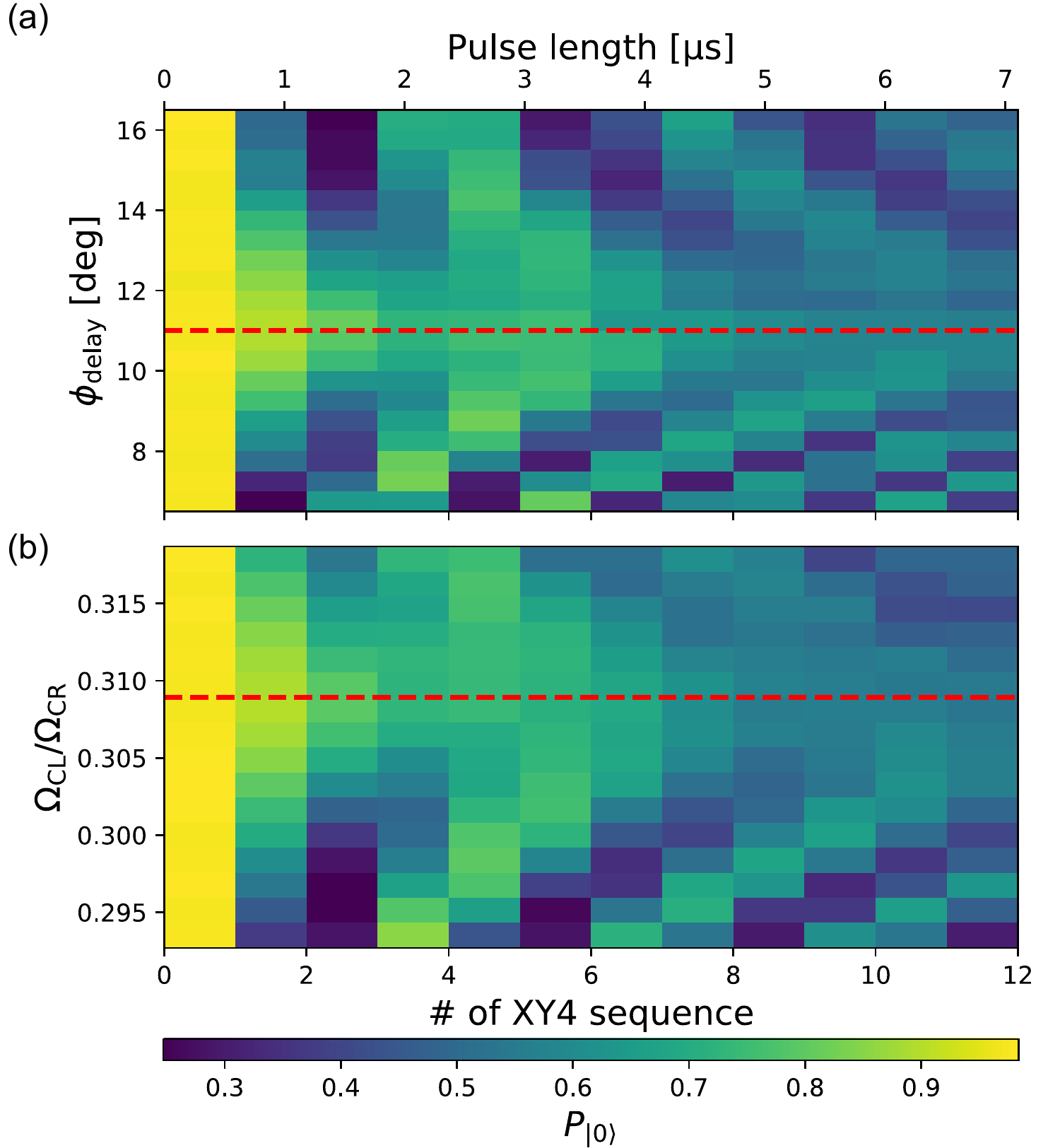}
    \caption{Ground state population of $Q_\unit{B}$ obtained by sweeping the XY4 sequence where $N_{xy4}=0,1,...,12$. (a) Result of sweeping $\phi_\unit{delay} = \phi_\unit{CL} - \phi_\unit{CR}$ with fixed $\phi_\unit{CR}$. (b) Result of sweeping $\Omega_\unit{CL}$ with fixed $\Omega_\unit{CR}$. Each red dashed lines show the initial guess minimizing $C_\unit{xy4}$.}
    \label{fig:xy4_sweep}
\end{figure}

First, to determine the amplitude $\Omega_\unit{CL}$ and phase $\phi_\unit{CL}$ of the CL drive, the amplitude $\Omega_\unit{CR}$ and phase $\phi_\unit{CR}$ of the CR drive were fixed, and the XY4 sequence was executed while changing the amplitude and phase of CL drive. The pulse sequence is
\begin{equation}
    \unit{XY4}_i = (Y_A \circ U_\unit{CR}^+ \circ X_A \circ U_\unit{CR}^+ \circ Y_A \circ U_\unit{CR}^+ \circ X_A \circ U_\unit{CR}^+)^i.
\end{equation}
Here, $i=0,1,...,N_{xy4}$, $U_\unit{CR}^\pm = U_\unit{CR}(\pm\Omega_\unit{CR},\phi_\unit{CR},\pm\Omega_\unit{CL},\phi_\unit{CL})$, and $X_A$ and $Y_A$ are $\pi$ pulse about the X and Y axes for $Q_\unit{A}$.
The XY4 sequence effectively cancels all the interactions that entangle the qubits, so that $Q_\unit{B}$ in the sequence evolves in time owing to the crosstalk terms $IX$ and $IY$, and if the parameters of the CL drive are correct, the state of $Q_\unit{B}$ does not change in time. First, as shown in Fig.~\ref{fig:xy4_sweep}(a, b),  $\phi_\unit{CL},\Omega_\unit{CL}$ was swept in order, and the parameters were determined so that the zig-zag pattern disappears and $C_\unit{xy4}=\sum_{i=0}^{N_{xy4}-1}|P_{0,i}^B-P_{0,i+1}^B|$ was minimized. Here, $P_{0,i}^B$ is the ground-state population of $Q_\unit{B}$ after $\unit{XY4}_i $ is applied.
As a result, the initial voltage ratio $m=\Omega_\unit{CL}/\Omega_\unit{CR}$ of the CR drive to the CL drive and the initial phase delay $\phi_\unit{delay} = \phi_\unit{CL} - \phi_\unit{CR}$ of the CL drive were estimated.

\begin{figure}
    \centering
    \includegraphics[width=8.6cm]{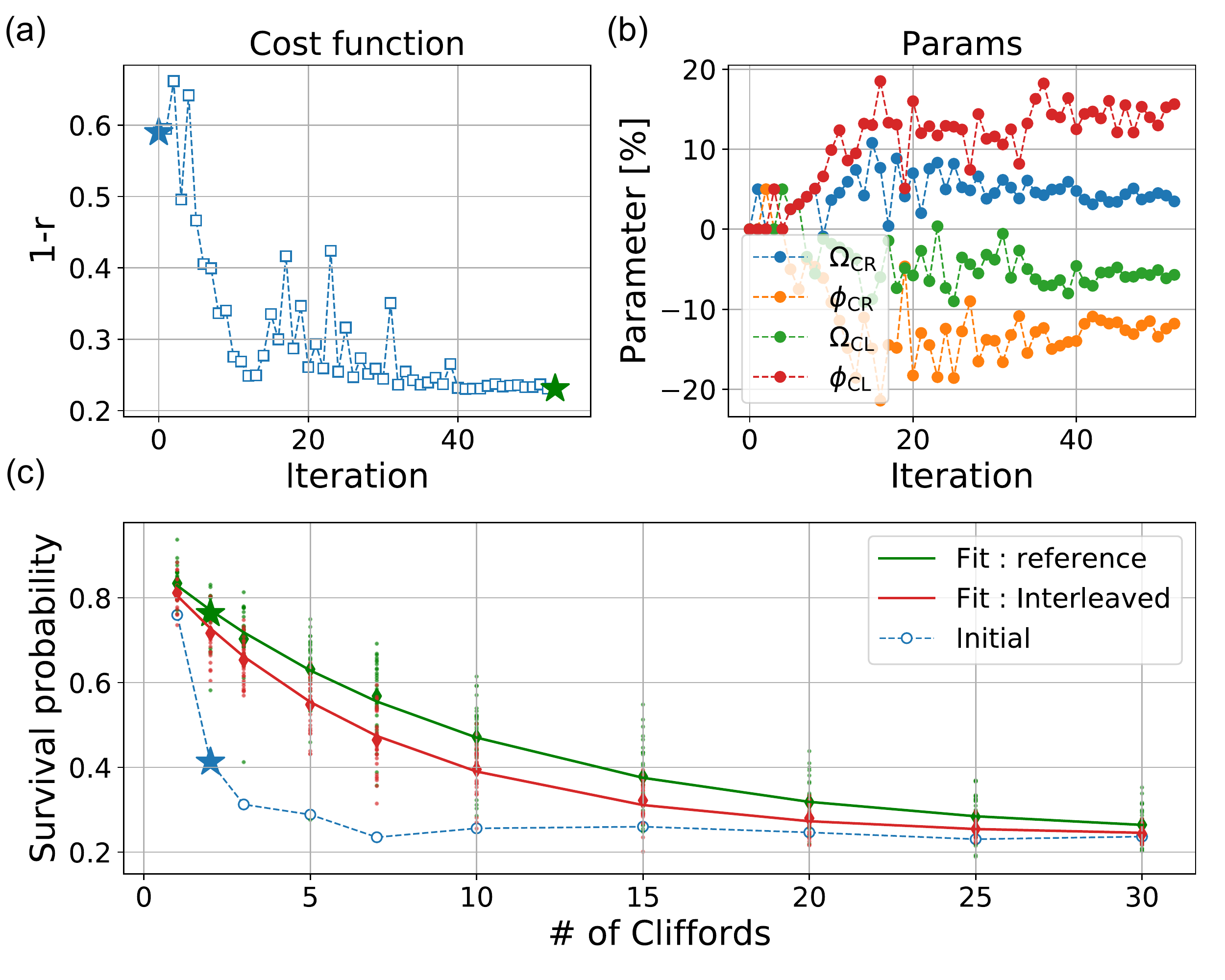}
    \caption{ECR-based CNOT gate calibration result obtained by ORBIT. (a) Cost function $1-r$, where $r \in [0,1]$ is the averaged survival probability of 20 RB sequences of gate length 2, plotted against the number of parameter updates. The blue star is the pre-optimization cost function, and the green star is the post-optimization cost function value. (b) Change in parameters from their initial values. (c) Results of the SRB and the IRB. The blue dashed line shows the result of the SRB before the ORBIT, and the green solid line shows the result of the reference SRB after the ORBIT. The red solid line shows the result of the IRB. The fidelity of the optimized CNOT gate was estimated to be 0.967(12) and the bracket shows 95\% confidence interval of the fitting result.}
    \label{fig:ORBIT_2Q_params}
\end{figure}

The initial $\Omega_\unit{CR},\,\phi_\unit{CR}$ were determined to be $U_\unit{CR}^+X_A U_\unit{CR}^- X_A \sim U_\unit{ZX}(+\pi/2)$ by fixing the values of $m$ and $\phi_\unit{delay}$ estimated above. Using the initial values determined from these results, the ORBIT was executed to obtain the high-fidelity CNOT gates. The Nelder-mead algorithm was used \cite{nelder1965simplex} and the cost function is the average of the survival probabilities of 20 RB sequences of Clifford length 2. The result of the ORBIT execution is shown in Fig.~\ref{fig:ORBIT_2Q_params}. We were able to find a better parameter set within approximately 20\% from the initial value.

\section{Reset calibration}
\begin{figure}
    \centering
    \includegraphics[width=17.8cm]{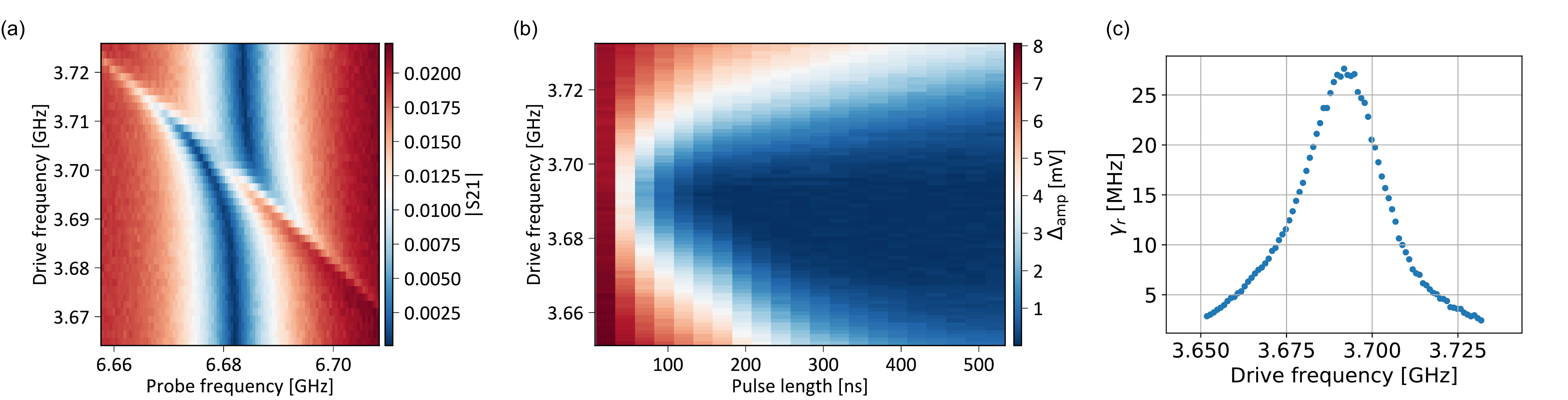}
    \caption{(a) Anti splitting by the effective interaction between $Q_\unit{B}$ and $R_\unit{B}$ during the $f0g1$ drive. The color shows the magnitude of the S-parameter (transmission). The vertical and horizontal axes represent the frequency of $f0g1$ drive applied to $Q_\unit{B}$ and applied to the $R_\unit{B}$, respectively. (b) Difference of the resonator response in the IQ plane between the response $(I', Q')$ after the reset sequence and the response $(I_0, Q_0)$ when $Q_\unit{B}$ is in the ground state, where $\Delta_\unit{amp}=\sqrt{(I'-I_0)^2 + (Q'-Q_0)^2}$. (c) Reset rate obtained by fitting (b).}
    \label{fig:Reset_summary}
\end{figure}

To generate the t-LCS, $Q_\unit{B}$ must be initialized because $\ket{\psi}_A\otimes \ket{0}_B$ must be input in the next cycle. In this work, we used the unconditional pulsed reset \cite{Pulsed_reset_protocol2018,PhysRevLett.121.060502}. The pulse sequence is shown in the main text. The reset sequence consists of $X_\pi^{ef}$ and $X_\pi^{f0g1}$ pulses with drive frequencies of $\omega_{ef}^\unit{B}/2\pi\simeq 5.053\,\unit{GHz}$ and $\omega_{f0g1}/2\pi\simeq 3.692\,\unit{GHz}$, respectively. For $X_\pi^{ef}$ pulse, the pulse waveform was Gaussian with a width of 32 ns. In addition, the amplitude was determined from the Rabi oscillation rate between the first and second excited states of $Q_\unit{B}$. And for $X_\pi^{f0g1}$ pulse, first we performed two-tone spectroscopy to estimate the drive frequency of $X_\pi^{f0g1}$ pulse as shown in Fig.~\ref{fig:Reset_summary}(a). Secondly, to determine the optimal drive frequency, $Q_\unit{B}$ was prepared in the second excited state, and the $f0g1$ pulse was applied while sweeping the pulse length and drive frequency and the result is shown in Fig.~\ref{fig:Reset_summary}(b). Using this result, we determined the optimal drive frequency by fitting it with $C_0e^{-\gamma_r t}+C_1$, where $\gamma_r$ and $C_{0,1}$ are reset rate and fitting constants, respectively. Finally, we got the optimal frequency $\omega_{f0g1}$. At this time, the drive power of $X_\pi^{f0g1}$ pulse was set to the maximum value in the available range of our setup, but since the resonator was strongly coupled to the environment, the oscillation was not visible and only the over-damping regime was accessible. Also, $X_\pi^{f0g1}$ pulse shape was flat-top Gaussian with a width of 260 ns and the total reset time is $300\,\unit{ns}$ with a buffer time of $8\,\unit{ns}$.  

\begin{figure}
    \centering
    \includegraphics[width=12cm]{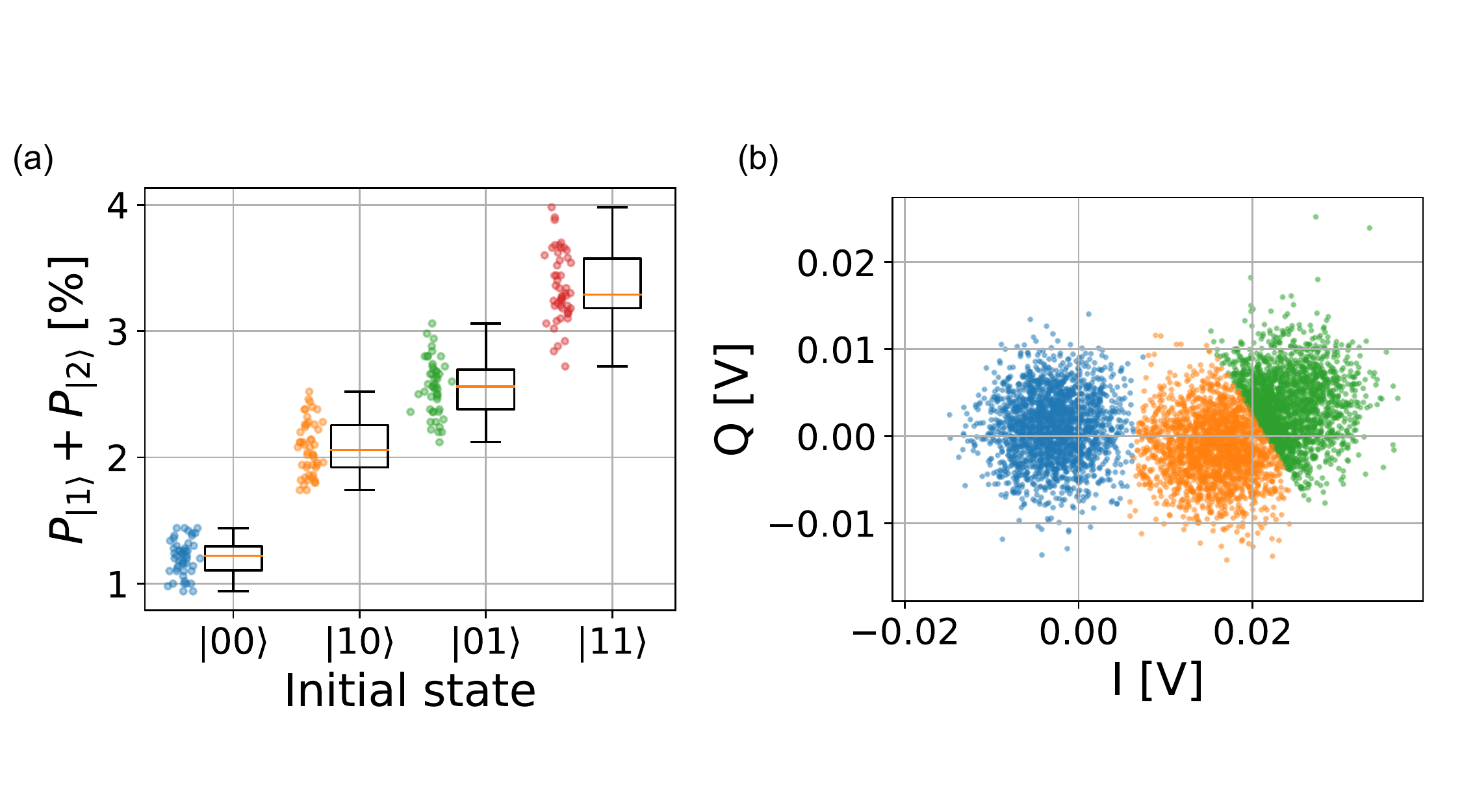}
    \caption{(a) Sum of populations remaining in the first and second excited states after the application of the reset pulse for different initial states. (b) Training result of the K-means filter used to determine the state from the digitized IQ data.}
    \label{fig:reset_data}
\end{figure}

Using the parameters determined above, the reset accuracy $F_{\ket{0}}$ was estimated from the average population remaining after the reset pulse [Fig.~\ref{fig:reset_data}(a)] with the initial state of the set to $\ket{00},\,\ket{01},\,\ket{10},$ or $\ket{11}$. For this estimation, we used a K-means filter \cite{Pulsed_reset_protocol2018} to classify the states and the training result is shown in Fig.~\ref{fig:reset_data}(b).

\begin{figure}
    \centering
    \includegraphics[width=8.6cm]{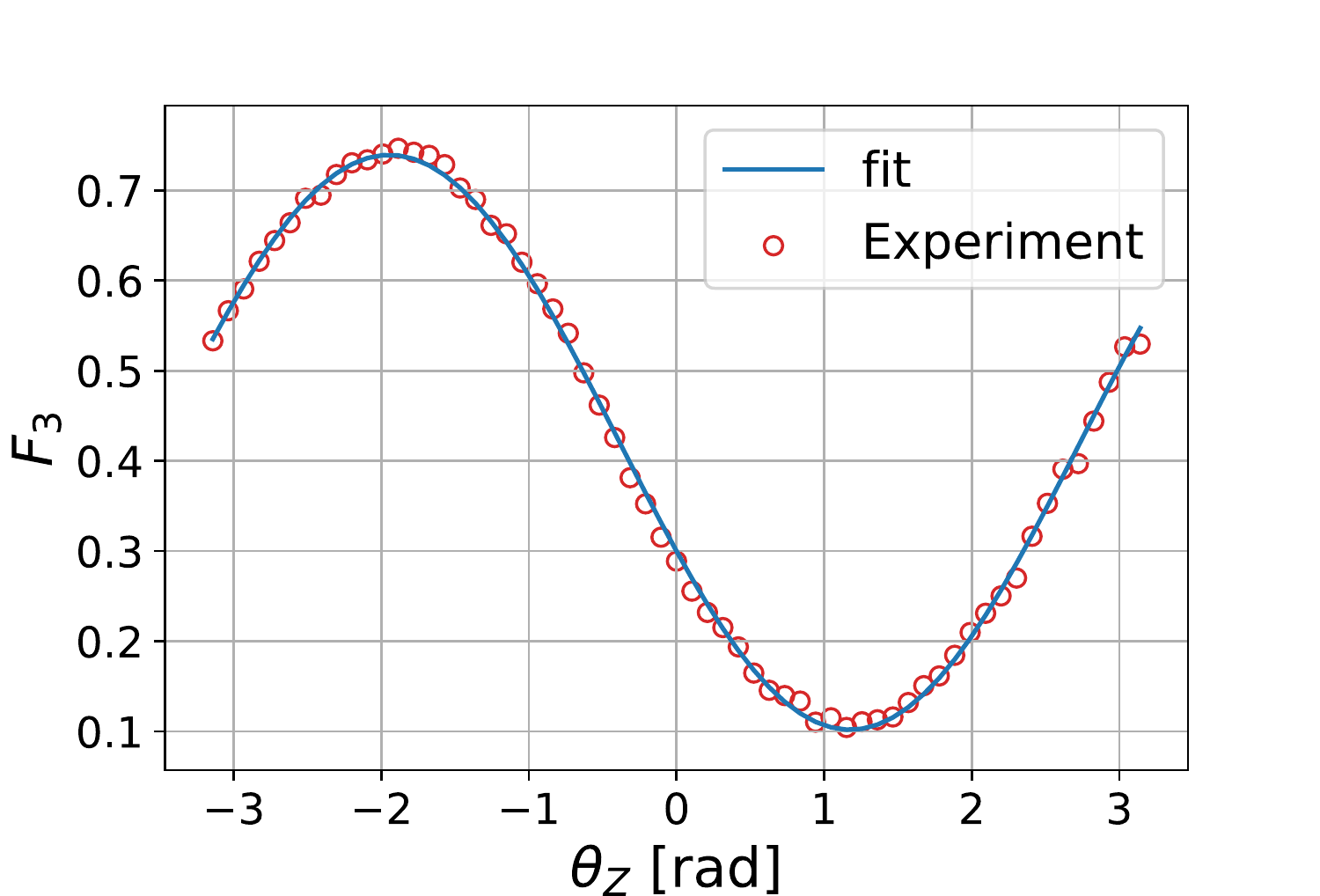}
    \caption{Result of quantum state tomography for three-qubit t-LCS while changing $\theta_Z$ and the fitting result by a sinusoidal function.}
    \label{fig:FQ3_rotZ_sweep}
\end{figure}

As mentioned in the main text, the $f0g1$ pulse has a large power which was about 15 dB larger than those of the other pulses. The $f0g1$ pulse induces an AC Stark shift of the $Q_\unit{A}$ frequency. Therefore, at the end of each cycle, a $\unit{Z}(\theta_Z)$ gate must be applied to correct the phase shift of $Q_\unit{A}$. To determine the amount of $\theta_Z$, state tomography was performed while changing the amount of rotation and fixing $N$ to $3$ (the number of repetitions is 2), and the obtained state fidelity $F_3(\theta_Z)$ is shown in Fig.~\ref{fig:FQ3_rotZ_sweep}. As a result of fitting the measurement result by a sinusoidal function, $\theta_Z^{opt}=-1.99\,\unit{rad}$ was determined.

\section{Numerical simulation method}
\begin{table*}
    \centering
    \begin{tabular}{|c|c|c|c|}\hline
        \multicolumn{2}{|c|}{Parameter name} & Value & Description \\ \hline
        $\unit{Q_A}$ & $T_1$              & 20 $\mathrm{\mu s}$ & Energy relaxation time \\ \cline{2-4}
                   & $T_2^*$            & 29 $\mathrm{\mu s}$ & Dephasing time \\ \cline{2-4}
                   & $t_{\pi/2}^A$ & 20 ns               & 1Q gate time ($\pi/2$ rotation) \\ \cline{2-4}
                   & $t_{\pi}^A$    & 40 ns               & 1Q gate time ($\pi$ rotation) \\ \cline{2-4}
                   & $t_\unit{CNOT}$ & 296 ns        & ECR gate + 1Q gate time                     \\ \cline{2-4}
                   & $t_b$        & 6 ns                & Buffer time between gates \\ \cline{2-4}
                   & $T_m^A$          & 750 ns              & Measurement time                           \\ \cline{2-4}
                   & $p_\unit{err}^m$ & 0.050                 & Measurement error probability \\ \hline\hline
        $\unit{Q_B}$ & $T_1$              & 20 $\mathrm{\mu s}$ & Energy relaxation time \\ \cline{2-4}
                   & $T_2^*$            & 27 $\mathrm{\mu s}$ & Dephasing time \\ \cline{2-4}
                   & $t_{\pi/2}^B$    & 20 ns               & 1Q gate time ($\pi/2$ rotation) \\ \cline{2-4}
                   & $t_{\pi}^B$      & 40 ns               & 1Q gate time ($\pi$ rotation)   \\ \cline{2-4}
                   & $t_b$  & 6 ns                & Buffer time between gates \\ \cline{2-4}
                   & $T_m^B$          & 750 ns              & Measurement time          \\ \cline{2-4}
                   & $T_r$          & 300 ns              & Reset time                         \\ \cline{2-4}
                   & $p_\unit{err}^m$ & 0.053                & Measurement error probability \\ \cline{2-4}
                   & $p_\unit{err}^\unit{ini}$ & 0.03                & Reset error probability    \\ \hline\hline
        Wait time  & $t_A$             & 1.45 $\mathrm{\mu s}$     & $T_m^B + T_r + t_m + t_r$ \\ \cline{2-4}
                   & $t_m$        & 0.2 $\mathrm{\mu s}$    & Wait time for resonator B to relax \\ \cline{2-4}
                   & $t_r$         & 0.2 $\mathrm{\mu s}$   & Wait time for resonator B to relax \\ \hline
    \end{tabular}
    \caption{Summary of parameters used in numerical simulation. For the coherence time, the value measured in the experiment was used. In the experiment, the wait time $t_m$ and $t_r$ are inserted after the readout pulse and after the reset pulse during the cycle, respectively.}
    \label{tab:params_sim}
\end{table*}

\emph{Gate and wait time errors}---For single-qubit gates, the CNOT gate and wait times in Tab.~\ref{tab:params_sim}, the accuracy of the coherence limit is assumed. The effects of $T_1$ and $T_2^*$ were taken into account as a following amplitude and phase damping error channel \cite{geller2013efficient}.
\begin{equation}
    \rho' = \Lambda_{APD}(\rho) = \sum_{m=1}^{3} K_m \rho K_m^\dagger,
\end{equation}
with Kraus matrices 
\begin{align}
    K_1 &= \left(
           \begin{matrix} 
               1 & 0 \\ 
               0 & \sqrt{1-\gamma-\lambda} 
           \end{matrix}
           \right), \\
    K_2 &= \left(
           \begin{matrix} 
               0 & \sqrt{\gamma} \\ 
               0 & 0 
           \end{matrix}
           \right), \\
    K_3 &= \left(
           \begin{matrix} 
               0 & 0 \\ 
               0 & \sqrt{\lambda} 
           \end{matrix}
           \right),
\end{align}
where
\begin{align}
    \gamma   &= 1 - \exp[-t_{g}/T_1], \\
    \lambda  &= \exp[-t_{g}/T_1] \left[ 1 - \exp[-2t_{g}/T_\phi] \right], \\
    T_\phi   &= \frac{2T_1T_2^*}{2T_1 - T_2^*}
\end{align}
and $t_g$ is the gate time.

\emph{Measurement error}---The readout errors for each qubit were modeld as following POVMs \cite{devitt2013quantum}:
\begin{align}
    F_0 &= (1-p_\unit{err}^m)\dyad{0}{0} + p_\unit{err}^m\dyad{1}{1}, \\
    F_1 &= (1-p_\unit{err}^m)\dyad{1}{1} + p_\unit{err}^m\dyad{0}{0},
\end{align}
where $p_\unit{err}^m = 1-F_a$ is the measurement error probability, also $F_0$ and $F_1$ satisfy $F_0+F_1=I$.
For state, $\rho$, the probability of obtaining the measured result 0 (1) is given by $\Tr[\rho F_0]\,(\Tr[\rho F_1])$. Furthermore, when the measured result $i\in\{0,1\}$ is obtained, the state is transformed as follows \cite{devitt2013quantum}:
\begin{equation}
    \rho \xrightarrow{i} \rho' = \frac{M_i \rho M_i^\dagger}{\Tr[\rho F_i]},
\end{equation}
where
\begin{align}
    M_0 &= \sqrt{1-p_\unit{err}^m}\dyad{0}{0} + \sqrt{p_\unit{err}^m}\dyad{1}{1}, \\
    M_1 &= \sqrt{1-p_\unit{err}^m}\dyad{1}{1} + \sqrt{p_\unit{err}^m}\dyad{0}{0}.
\end{align}

\emph{Reset error}---For the reset error of $Q_\unit{B}$, we assume $\rho$ is the two qubit state. To model the qubit reset, $Q_\unit{B}$ is traced out firstly and replaced with a new $\dyad{0}{0}_\unit{B}$ and we obtain a state $\rho'=\Tr_\unit{B}[\rho]\otimes\dyad{0}{0}_\unit{B}$. After this ideal reset of $Q_\unit{B}$, the following stochastic bit flip channel is applied to $\rho'$
\begin{equation}
    \mathcal{E}(\rho') = (1-p_\unit{err}^{ini})\rho' + p_\unit{err}^{ini} X \rho' X^\dagger,
\end{equation}
where $X$ is a not gate and acts on $Q_\unit{B}$, and $p_\unit{err}^{ini} = 1-F_{\ket{0}}$ is the reset error probability.
\end{document}